%% file: mass-main.tex
\documentclass[10pt,conference,letterpaper]{IEEEtran}

\IEEEoverridecommandlockouts

\usepackage[T1]{fontenc}
\usepackage{cite}
\usepackage{import}
\usepackage[caption=false]{subfig}
\usepackage{graphicx}
\usepackage{balance}
\usepackage{tikz}
\usepackage{textcomp}
\usepackage{listings}
\usepackage{makecell}
\usepackage{multirow}
\usepackage{rotating}
\usepackage{diagbox}
\usepackage{xurl}
\usepackage{amssymb}
\usepackage{xcolor}
\usepackage[absolute,showboxes]{textpos}
\usepackage{svg}
\usepackage{soul}
\usepackage{multicol}
\usepackage[linesnumbered,ruled,vlined]{algorithm2e}
\usepackage[cmex10]{amsmath}
\usepackage{cleveref}
\usepackage{amsthm}
\usepackage{amsfonts}
\usepackage{mathbbol}
\usepackage{bm} 
\newtheorem{theorem}{Theorem}
\newtheorem{lemma}[theorem]{Lemma}
\newtheorem{result}[theorem]{Result}
\newtheorem{definition}[theorem]{Definition}

\usepackage[none]{hyphenat}

\begin{document}
	
	\input{header}
	\maketitle

	\input{abstract}
	\input{introduction}
	\input{system}
	\input{formulation}
	\input{heuristic}
	\input{results}
 	\input{related_work}
	\input{conclusions}

	\bibliographystyle{IEEEtran}
	\bibliography{IEEEabrv,mass-main.bib}

\end{document}

%% file: header.tex
%
\title{Joint UPF and Edge Applications Placement and Routing in 5G \& Beyond 
}

\author{
 	\IEEEauthorblockN{Endri Goshi\IEEEauthorrefmark{2}, Hasanin Harkous\IEEEauthorrefmark{1}, Shohreh Ahvar\IEEEauthorrefmark{3}, Rastin Pries\IEEEauthorrefmark{1}, Fidan Mehmeti\IEEEauthorrefmark{2}, Wolfgang Kellerer\IEEEauthorrefmark{2}}
 	
 	\IEEEauthorrefmark{2}Technical University of Munich, \IEEEauthorrefmark{1}Nokia Germany, \IEEEauthorrefmark{3}Nokia France\\
 	
 	E-Mail:\IEEEauthorrefmark{2}\{endri.goshi, fidan.mehmeti, wolfgang.kellerer\}@tum.de,\\ \IEEEauthorrefmark{1}\{hasanin.harkous, shohreh.ahvar, rastin.pries\}@nokia.com, 

}

%% file: abstract.tex
\begin{abstract}
The development of 5G networks has enabled support for a vast number of applications with stringent traffic requirements, both in terms of communication and computation. Furthermore, the proximity of the entities, such as edge servers and User Plane Functions (UPFs) that provide these resources is of paramount importance.
However, with the ever-increasing demand from these applications, operators often find their resources insufficient to accommodate all requests.
Some of these demands can be forwarded to external entities, not owned by the operator.
This introduces a cost, reducing the operator's profit.
Hence, to maximize operator's profit, it is important to place the demands optimally in internal or external edge nodes.
To this end, we formulate a constrained optimization problem that captures this objective and the inter-play between different parameters, which turns out to be NP-hard.
Therefore, we resort to proposing a heuristic algorithm which ranks the demands according to their value to the operator and amount of resources they need. Results show that our approach outperforms the benchmark algorithms, deviating from the optimal solution by only $\sim3\%$ on average.
\end{abstract}

\begin{IEEEkeywords}
5G core, UPF, Edge clouds, Routing. 
\end{IEEEkeywords}

%% file: introduction.tex
\section{Introduction}
\label{sec:introduction}
The introduction and deployment of 5G networks has enabled an array of new use-cases, and applications that leverage the high-throughput, scalability and flexibility of such networks.
Computationally-intensive and delay-sensitive applications, such as Augmented and extended Reality (AR/XR), rely on the availability of the compute and network resources at users' proximity.
Through technologies such as Network Function Virtualization (NFV), network resources can be leveraged by flexibly deploying User Plane Functions (UPFs)\cite{blanco}.
Moreover, Multi-Access Edge Computing (MEC) enables the users to offload their compute tasks to servers deployed on the edge of the network.

Currently, MEC solutions consider a separate orchestration of the compute and network resources, where cloud operators and Telco Network Operators (TNOs) manage only the resources of their domain.
The importance of the joint consideration of these types of resources is highlighted by concepts such as in-network computing and Compute and Network Convergence (CNC)\cite{cnc}.
Collaboration initiatives, such as AWS Wavelength\footnote{https://aws.amazon.com/wavelength} and Verizon Edge Discovery Service (EDS) \footnote{https://www.verizon.com/business/5g-edge-portal}, aim to facilitate the orchestration of Edge Applications (EAs) close to the TNO's UPFs.
These approaches have two main shortcomings: i) EA orchestration burden falls on the EA providers, who must know their users’ locations to orchestrate compute resources efficiently; and ii) TNOs lack prior information about the deployment locations of specific EAs, resulting in inefficient orchestration of network resources.

Therefore, in this work, we propose a different approach where the TNOs establish agreements with EA providers to take over the orchestration of the EAs, in addition to the UPF.
This way, they can efficiently route user demands within the transport network, while establishing a new line of revenue through the deals made with the EA providers.
Moreover, the same approach is applicable to private/campus 5G network operators that want to expand the services they offer in their network and efficiently utilize their edge infrastructure.

Research works that jointly investigate the allocation of compute and network resources, consider the problem from a network planning perspective.
As such, they focus on optimally building and dimensioning the edge infrastructure and the deployment of UPFs, predominately assuming a $1$-to-$1$ association between the Base Stations (BSs), edge nodes, and UPFs \cite{Li2021, Li2022, Zhang2022, Wang2019}.
In contrast, in this work, we consider an already-deployed edge infrastructure and transport network.
A holistic approach to the problem is taken, integrating aspects such as UPF and EA deployment, user demand placement and routing, in a single problem formulation.
Modeling such a system with realistic assumptions presents challenging tasks, especially given the multiple dimensions of the problem.

To this end, we model a system with integrated compute and network infrastructure, and then formulate the joint UPF and EA placement and routing problem using Integer Linear Programming (ILP).
We show that the problem is NP-hard, and therefore, we design and implement a greedy heuristic called \emph{RanGr}.
Our evaluations show that \emph{RanGr} is able to find high-quality solutions within a very short time, making it a viable algorithm to be used by TNOs for the orchestration of their compute and network resources.

Specifically, our main contributions are:
\begin{itemize}
\item We formulate a constrained optimization problem that captures reliably the system's behavior and the interplay between different components and resources.

\item We show that the problem is NP-hard and propose a polynomial-time algorithm which provides near-optimal performance. 
\item Conducting extensive realistic simulations, we show that our approach outperforms the benchmark algorithms significantly in all considered scenarios.
\end{itemize}

The remainder of the paper is organized as follows.
A detailed explanation of the system model is provided in Section~\ref{sec:system_design}, followed by the problem formulation in Section~\ref{sec:prob_form}.
Section~\ref{sec:rangr} introduces \emph{RanGr}, our proposed heuristic approach, while Section~\ref{sec:results} details the evaluation scenarios and the obtained results.
Section~\ref{sec:related_work} presents related work in the areas of edge infrastructure and UPF deployment.
Lastly, Section~\ref{sec:conclusions} discusses future work and concludes the paper.

%% file: system.tex
\section{System Model}
\label{sec:system_design}

\begin{figure*}[t]
	\centering
	\includegraphics[width=6.2in]{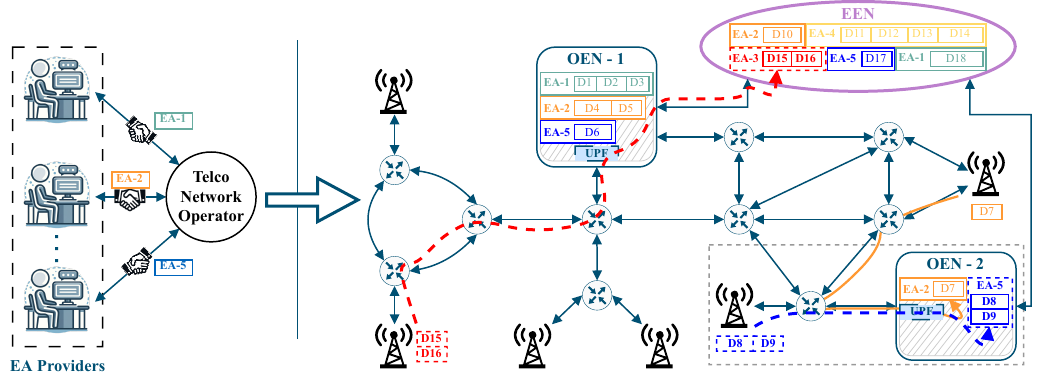}
	\caption{System overview depicting the initialization of UPFs and EAs, and the placement of demands in the OENs and the EEN.
	As resources in OENs become scarce, demands are only anchored in the UPFs and are then offloaded to the EEN.
	} 
	\label{fig:system}
\end{figure*}

An overview of the system considered in this work is given in Fig.~\ref{fig:system}.
The edge infrastructure consists of two types of edge nodes: i) Operator Edge Node (OEN), and ii) External Edge Node (EEN).
The former, as the name suggests, represents edge servers owned by TNO.
Being within the operator's domain, they are used to deploy the UPFs, EAs, and place incoming demands to be served by the EAs.
OENs are denoted by $e \in \mathcal{E}$ and there are $\vert\mathcal{E}\vert$ OENs comprising the edge infrastructure.
Each OEN $e$ is characterized by its processing and storage capacity, denoted by $P_e$ and $S_e$, respectively, and by its operational cost $C_e$.
On the other hand, the EEN, denoted by $e^*$, is an abstraction of edge servers not owned by the operator (e.g., provided by AWS or Google).
The resources in these nodes, which are assumed to be abundant, are rented by operators for additional deployment of EAs and demands.

In addition, the edge network comprises $\vert\mathcal{J}\vert$ BSs denoted by $j \in \mathcal{J}$, $\vert\mathcal{K}\vert$ forwarding nodes (i.e., switches, routers, etc.), denoted by $k \in \mathcal{K}$, and $\vert\mathcal{V}\vert$ links interconnecting the transport network, denoted by $v \in \mathcal{V}$.
Each link $v$ is characterized by its bandwidth capacity $B_v$, and the link latency $t^v$\footnote{Link latency is defined as the sum of the transport and propagation latencies.
We assume that link latency between any OEN and EEN is higher than the link latency between any two nodes in operator's transport network.}.
In this work, we assume that the forwarding nodes can process the incoming traffic at line rate, as long as the link capacity is not exceeded.
Thus, we do not model their capacity and processing latency.
The edge network topology is modeled as a directed graph $G=(\mathcal{J} \cup \mathcal{K} \cup \mathcal{E} \cup e^*, \mathcal{V})$.

\subsection{EAs and User Demands}
In our system, agreements exist between EA providers and TNOs to manage the orchestration of $\vert\mathcal{I}\vert$ different EAs denoted by $i \in \mathcal{I}$.
They represents services (e.g., AR/XR applications) that are deployed on edge nodes and serve the incoming user demands.
EAs are characterized by their idle CPU utilization $p_i$ (e.g., due to monitoring) and storage requirement $s_i$.

Demands are abstracted as Packet Data Unit (PDU) sessions.
In this work, we consider the BS as the source of a demand since that is the entry point of the traffic into the transport network, and thus, demands can belong to the same user or to different ones, making the approach general enough.
From each BS $j$ there are $\vert\mathcal{L}_{j,i}\vert$ demands of EA type $i$ that need to be served, denoted by $l \in \mathcal{L}_{j,i}$.
Demands are uniquely identified by the tuple $\{j,i,l\}$, and are characterized by their CPU requirement $p_{j,i,l}$, bandwidth requirement $b_{j,i,l}$, and the maximum delay budget $t_{j,i,l}$.
Moreover, serving a demand (i.e., placing it on an OEN or EEN) generates revenue for the operator.
This revenue is captured by the utility value, which is denoted as $U_{j,i,l}$.
Conversely, the cost for renting resources in the EEN to place a demand is captured by $C_{j,i,l}$.

\begin{table}[t]
	\centering
	\caption{Summary of notations used in this paper }
	\begin{tabular}{|l|l|}
		\hline
		\textbf{Notation}       									&	\textbf{Description}           							\\ \hline
		$j \in \mathcal{J}:\{1, 2, ..., |\mathcal{J}|\} $						&	Base Stations (BS)										\\ \hline
		
		$i \in \mathcal{I}:\{1, 2, ..., |\mathcal{I}|\} $						&	Types of Edge Applications (EAs)								\\ \hline
		
		$l \in \mathcal{L}_{j,i}:\{1, 2, ..., |\mathcal{L}_{j,i}|\}$					&	Demands originating from BS~$j$ 	  					\\ \hline
		
		$e \in \mathcal{E}:\{1, 2, ..., |\mathcal{E}|\} $						&	Operator Edge Nodes (OENs)								\\ \hline
		
		$e^*$														&	External Edge Node (EEN)								\\ \hline
		
		$k \in \mathcal{K}:\{1, 2, ..., |\mathcal{K}|\} $						&	Forwarding nodes in the topology									\\ \hline
		
		$v \in \mathcal{V}:\{1, 2, ..., |\mathcal{V}|\} $						&	Links in the topology									\\ \hline
		
		$n \in \mathcal{N}:\{1, 2, ..., |\mathcal{N}|\} $						&	Paths from BSs to OENs and EEN							\\ \hline
		
		$r \in \mathcal{R}:\{1, 2, ..., |\mathcal{R}|\} $						&	UPF replicas than can be deployed						\\ \hline
		
		$U_{i,j,l}$													&	Utility of serving demand~$\{j,i,l\}$					\\ \hline
		$C_{j,i,l}$													&	{\begin{tabular}[l]{@{}l@{}}
				Cost of offloading demand~$\{j,i,l\}$ to EEN
		\end{tabular}}											\\ \hline
		$C_e$														&	Cost of switching on OEN~$e$							\\ \hline
		
		$P_e$														& 	Total CPU capacity of OEN~$e$ 			 				\\ \hline
		
		$p_i$														& 	Idle CPU utilization of EA~$i$							\\ \hline
		
		$p_{j,i,l}$													& 	CPU requirement of demand~$\{j,i,l\}$ 					\\ \hline
		
		$p_{r}$													& 	CPU requirement of $r$~UPF replicas 					\\ \hline
		
		$S_e$														& 	Storage capacity of OEN~$e$		 						\\ \hline
		
		$s_i$														& 	Storage requirement of EA~$i$ 							\\ \hline
		
		$B_v$														& 	Bandwidth capacity of link~$v$							\\ \hline
		
		$B_{r}$													&	Bandwidth capacity of $r$~UPF replicas					\\ \hline
		
		$b_{j,i,l}$													& 	Bandwidth requirement of demand~$\{j,i,l\}$				\\ \hline
		
		$t_{j,i,l}$									& 	Maximum latency  of demand~$\{j,i,l\}$ 			\\ \hline
		
		$t^{v}$														& 	Latency of link $v$				\\ \hline

		$z_e \in \{0, 1\}$											&	Whether OEN~$e$	is switched on							\\ \hline
		
		$d_{j,i,l,e} \in \{0, 1\}$									& 	{\begin{tabular}[l]{@{}l@{}}
				If demand~$\{j,i,l\}$ is served at OEN~$e$
		\end{tabular}}											\\ \hline
		
		$T_{j,i,l} \in \{0, 1\}$									& 	{\begin{tabular}[l]{@{}l@{}}
				If demand~$\{j,i,l\}$ is served at EEN
		\end{tabular}}											\\ \hline
		
		$F_{j,i,l} \in \{0, 1\}$									& 	{\begin{tabular}[l]{@{}l@{}}
				Whether demand~$\{j,i,l\}$ is rejected
		\end{tabular}}											\\ \hline
		
		$x_{i,e} \in \{0, 1\}$										& 	{\begin{tabular}[l]{@{}l@{}}
				Whether EA~$i$ is initialized on OEN~$e$
		\end{tabular}}											\\ \hline
		
		$w_{e}^{r} \in \{0, 1\}$									& 	{\begin{tabular}[l]{@{}l@{}}
				If $r$~UPF replicas deployed on OEN~$e$
		\end{tabular}}											\\ \hline
		
		$w_{j,i,l,e} \in \{0, 1\}$									& 	{\begin{tabular}[l]{@{}l@{}}
				Whether demand~$\{j,i,l\}$ is anchored				\\
				in a UPF deployed on OEN~$e$
		\end{tabular}}											\\ \hline
		
		$\theta_{j,i,l}^{n} \in \{0, 1\}$							& 	{\begin{tabular}[l]{@{}l@{}}
				Whether path~$n$ is chosen to route demand			\\
				$\{j,i,l\}$ to the OEN or EEN
		\end{tabular}}											\\ \hline
	
		$\delta_{j,i,l}^{n,v} \in \{0, 1\}$						& 	{\begin{tabular}[l]{@{}l@{}}
				Whether link~$v$ is in path~$n$ that routes			\\
				demand~$\{j,i,l\}$ to the OEN or EEN
		\end{tabular}}											\\ \hline		
		
	\end{tabular}
 \label{tab:notation}
\end{table}

\subsection{UPFs and Routing}
One of the new key aspects of 5G Core architecture is the softwarization of its entities, which was enabled by advancements in NFV and programmable data planes technologies~\cite{p4upf}.
Therefore, UPFs are now commonly deployed as Virtual Network Functions (VNFs) on general-purpose servers.
On this basis, UPFs in our system are modeled as VNFs. 
Clusters of UPFs are dynamically initialized on OENs and can be \emph{horizontally scaled} (i.e., new replicas are deployed to keep up with the input traffic), where $r \in \mathcal{R}$ denotes the number of UPFs in a cluster, and $R=\vert\mathcal{R}\vert$ is the maximum cluster size.
For a UPF of scale $r$ deployed on an OEN, $p_{r}$ CPU resources have to be reserved, amounting to a total processing bandwidth of $B_{r}$.

One of prime tasks accomplished by UPF is to anchor the user traffic that leaves the core network, and then to forward it to its destination.
In our system, the user traffic is represented by the demands, whose destinations are the EAs deployed on OENs or EEN.
Therefore, ``accepting'' a demand means not only finding a suitable server with enough resources to place the demand, but also finding a viable path from the source BS to an anchoring UPF, and further to the destination EA.
We assume that if a demand is to be placed on an OEN, it must be anchored on a UPF instance deployed on that same OEN.
To illustrate this with an example from Fig.~\ref{fig:system}, demands D$1$-D$6$ are anchored on the UPF in OEN-$1$.
Otherwise, if the demand is to be offloaded to the EEN (e.g., demands D$15$ and D$16$ shown in red in Fig.~\ref{fig:system}), it can be anchored on a UPF instance deployed on any viable OEN.
To this end, we rely on a \emph{link-path} formulation~\cite{link-path-book}, where a set of $\vert\mathcal{N}\vert$ paths from any BS to any OEN and the EEN is built, and each path is denoted by $n \in \mathcal{N}$.

Table~\ref{tab:notation} summarizes the notation used throughout this paper. 

%% file: formulation.tex
\section{Problem Formulation}
\label{sec:prob_form}
In this section, we present the ILP formulation of the optimization problem, and then show that it is a variation of the Knapsack Problem and therefore is NP-hard. 

\subsection{ILP Formulation}
\label{subsec:ilp_formulation}

In this work, we focus on maximizing the utility derived from demands that are accepted and served on the OENs. We also consider the costs associated with keeping the OENs operational and the rental costs incurred from offloading demands to the EEN. The objective is to maximize the following function:
\begin{align}
	\label{obj}
	\max\left({\sum_{j,i,l}(U_{j,i,l} \hspace{-2pt}\cdot \hspace{-2pt}(1 \hspace{-2pt}- \hspace{-2pt}F_{j,i,l}) \hspace{-2pt}- \hspace{-2pt}C_{j,i,l} \hspace{-2pt}\cdot \hspace{-2pt}T_{j,i,l}) \hspace{-2pt}- \hspace{-2pt}\sum_{e} z_e \hspace{-2pt}\cdot \hspace{-2pt}C_e}\right),
\end{align}
where $U_{j,i,l}$ represents the utility gain from accepting a demand, $F_{j,i,l}\in\{0,1\}$ denotes whether demand $\{j,i,l\}$ is rejected, $C_{j,i,l}$ denotes costs incurred if demand is placed on EEN, and $T_{j,i,l}\in\{0,1\}$ denotes whether the demand is offloaded to the EEN.
$z_e\in\{0,1\}$ denotes whether OEN $e$ is operational and $C_e$ denotes the operational costs.
The first term refers to the utility gained from accepting a demand, the second one refers to the cost paid to the EEN operator, and the last term captures the operating cost of OENs.

Before describing the constraints, it should be mentioned that for routing we consider a \emph{link-path} formulation~\cite{link-path-book}, aimed to find the best path from source BS to destination OEN/EEN.
Link-path formulation allows for a more granular representation of routes and captures constraints and objectives related to path characteristics (e.g., latency, bandwidth) more effectively.
In total, there are $17$ constraints related to the finite resources of OENs, UPFs, network resources, EA deployment and demand anchoring and placement. 

\noindent \textit{1) OEN capacity constraints:}
The CPU resources allocated for the deployment of UPFs, initialization of EAs, and placement of demands, should not exceed the total CPU capacity of OEN:
\begin{equation}
\label{const:first}
	\sum_{r=1}^{|\mathcal{R}|}p_{r} w_{r,e} \hspace{-3pt}+ \hspace{-3pt}\sum_{i=1}^{|\mathcal{I}|} p_i x_{i,e} \hspace{-3pt}+\hspace{-3pt} \sum_{j=1}^{|\mathcal{J}|} \sum_{i=1}^{|\mathcal{I}|} \sum_{l=1}^{|\mathcal{L}_{j,i}|} p_{j,i,l} d_{j,i,l,e} \le P_e, 
	\forall e \in \mathcal{E},
\end{equation}
where $w_{r,e}\in\{0,1\}$ denotes whether a UPF cluster with $r$ replicas is initialized on OEN $e$, $x_{i,e}\in\{0,1\}$ denotes whether EA $i$ is deployed on OEN $e$, and $d_{j,i,l,e}\in\{0,1\}$ denotes whether demand $\{j,i,l\}$ is placed on OEN $e$.

The storage resources taken up by the deployment of EAs should not exceed the total storage capacity of the OEN:
\begin{align}
	\sum_{i=1}^{|\mathcal{I}|} s_i \cdot x_{i,e} \le S_e, \quad \forall e \in \mathcal{E}.
\end{align}

\noindent \textit{2) EA deployment constraint:}
EAs can be deployed on an OEN only if the OEN is in operational mode (\textit{ON}):
\begin{align}
	x_{i,e} \le z_{e}, \quad \forall i \in \mathcal{I}, \forall e \in \mathcal{E}.
\end{align}

\noindent \textit{3) UPF deployment constraints:}
UPF clusters can be initialized in OENs only if OEN is in operational mode (\textit{ON}):
\begin{align}
	\sum\limits_{r=1}^{|\mathcal{R}|}w_{r,e} \le z_{e}, \quad \forall e \in \mathcal{E}.
\end{align}

On an OEN, there can only be one UPF cluster deployed:
\begin{align}
	\sum\limits_{r=1}^{|\mathcal{R}|}w_{r,e} \le 1, \quad \forall e \in \mathcal{E}.
\end{align}

\noindent \textit{4) Demand placement constraints:}
A demand can be deployed on an OEN only if a same type EA is initialized on OEN:
\begin{align}
	d_{j,i,l,e} \le x_{i,e}, \quad \forall j \in \mathcal{J}, \forall i \in \mathcal{I}, \forall l \in \mathcal{L}_{j,i}, \forall e \in \mathcal{E}.
\end{align}

An EA should be initialized on an OEN only if there are demands of the same type that will be placed on the OEN:
\begin{align}
	\sum_{j=1}^{|\mathcal{J}|}\sum_{l=1}^{|\mathcal{L}_{j,i}|} d_{j,i,l,e} \ge x_{i,e}, \quad \forall i \in \mathcal{I}, \forall e \in \mathcal{E}.
\end{align}

There are three options for handling a demand: it can be rejected, offloaded to the EEN, or placed on exactly one OEN:
\begin{equation}
	F_{j,i,l} + T_{j,i,l} + \sum_{e=1}^{|\mathcal{E}|}d_{j,i,l,e} = 1, \ \forall j \in \mathcal{J}, \forall i \in \mathcal{I}, \forall l \in \mathcal{L}_{j,i}.
\end{equation}

\noindent \textit{5) Demand anchoring constraints:}
An accepted demand must be anchored on exactly one OEN:
\begin{align}
	F_{j,i,l} + \sum\limits_{e=1}^{|\mathcal{E}|}w_{j,i,l,e} = 1, \quad \forall j \in \mathcal{J}, \forall i \in \mathcal{I}, \forall l \in \mathcal{L}_{j,i},
\end{align}
where $w_{j,i,l,e}\in\{0,1\}$ denotes whether demand $\{j,i,l\}$ is anchored on the UPF cluster deployed on OEN $e$.

Demands will be anchored on OEN $e$ if they are also placed on OEN $e$, or placed on the EEN but OEN $e$ is the last hop in the path:
\vspace{-3pt}
\begin{align}
	&w_{j,i,l,e} = d_{j,i,l,e} + \sum_{n=1}^{|\mathcal{N}|}\delta_{j,i,l}^{n,v}, \\
	&\quad \forall j \in \mathcal{J}, \forall i \in \mathcal{I}, \forall l \in \mathcal{L}_{j,i}, \forall e \in \mathcal{E}, v\hspace{-2pt}=\hspace{-2pt}\{e, e^*\},\nonumber
\end{align}
where $\delta_{j,i,l}^{n,v}\in\{0,1\}$ denotes whether link $v$ is part of the path $n$ that routes the demand $\{j,i,l\}$. As $v\hspace{-2pt}=\hspace{-2pt}\{e, e^*\}$, the constraint considers only the last hop from any OEN to EEN.

A demand can be anchored on an OEN, if there is a UPF cluster deployed on that OEN:
\begin{align}
	w_{j,i,l,e}\hspace{-2pt}\le\hspace{-2pt}\sum_{r=1}^{|\mathcal{R}|}w_{r,e}, \quad \forall j \in \mathcal{J}, \forall i \in \mathcal{I}, \forall l \in \mathcal{L}_{j,i}, \forall e \in \mathcal{E}.
\end{align}

UPF clusters are initialized on an OEN only if there are demands anchored on that OEN:
\begin{align}
	\sum_{j=1}^{|\mathcal{J}|} \sum_{i=1}^{|\mathcal{I}|} \sum_{l=1}^{|\mathcal{L}_{j,i}|} w_{j,i,l,e} \ge \sum_{r=1}^{|\mathcal{R}|}w_{r,e}, \quad \forall e \in \mathcal{E}.
\end{align}

\noindent \textit{6) Routing constraints:}
Accepted demands can be routed through exactly one path:
\begin{align}
	\sum\limits_{n=1}^{|\mathcal{N}|} \theta_{j,i,l}^{n}= d_{j,i,l,e} &+ T_{j,i,l}, \\
	&\forall j \in \mathcal{J}, \forall i \in \mathcal{I}, \forall l \in \mathcal{L}_{j,i}, \forall e \in \mathcal{E}, \nonumber
\end{align}
where $\theta_{j,i,l}^{n} \in \{0,1\}$ denotes whether path $n$ is used to route demand $\{j,i,l\}$.

If a demand is routed through a path, then all the links in that path are used to transport the demand:
\begin{equation}
	\hspace{-6pt}\delta_{j,i,l}^{n,v} \hspace{-3pt}=\hspace{-3pt} \theta_{j,i,l}^{n}, 
	\ \forall j \in \mathcal{J}, \forall i \in \mathcal{I}, \forall l \in \mathcal{L}_{j,i}, \forall n \in \mathcal{N}, \forall v \in \mathcal{V}. 
\end{equation}

The latency incurred in the selected path (as a function of the number of links in the path and the links' latency) cannot exceed the maximum delay budget for the demand:
\begin{align}
	\hspace{-6pt}\sum_{v=1}^{|\mathcal{V}|}\delta_{j,i,l}^{n,v} \cdot t^{v} \le t_{j,i,l}, \hspace{1pt} \forall j \in \mathcal{J}, \forall i \in \mathcal{I}, \forall l \in \mathcal{L}_{j,i}, \forall n \in \mathcal{N}.
\end{align}

\noindent \textit{7) Network resources:}
The total volume of traffic that is anchored on an OEN cannot exceed the total capacity of OEN's UPF cluster:
\begin{align}
	\sum_{j=1}^{|\mathcal{J}|} \sum_{i=1}^{|\mathcal{I}|} \sum_{l=1}^{|\mathcal{L}_{j,i}|} w_{j,i,l,e} \cdot b_{j,i,l} \le \sum_{r=1}^{|\mathcal{R}|}w_{r,e} \cdot B_{r}, \quad \forall e \in \mathcal{E}.
\end{align}

The total volume of traffic transported through a link cannot exceed the bandwidth of the link:
\begin{align}
	\label{const:last}
	\sum_{j=1}^{|\mathcal{J}|} \sum_{i=1}^{|\mathcal{I}|} \sum_{l=1}^{|\mathcal{L}_{j,i}|} \sum_{n=1}^{|\mathcal{N}|} \delta_{j,i,l}^{n,v} \cdot b_{j,i,l} \le B_v, \quad \forall v \in \mathcal{V}.
\end{align}	

To summarize, the constrained optimization problem related to our system, and in the focus of this paper, is:
\begin{align*}
	\max\left({\sum_{j,i,l}(U_{j,i,l} \hspace{-2pt}\cdot \hspace{-2pt}(1 \hspace{-2pt}- \hspace{-2pt}F_{j,i,l}) \hspace{-2pt}- \hspace{-2pt}C_{j,i,l} \hspace{-2pt}\cdot \hspace{-2pt}T_{j,i,l}) \hspace{-2pt}- \hspace{-2pt}\sum_{e} z_e \hspace{-2pt}\cdot \hspace{-2pt}C_e}\right)
\end{align*}
\vspace{-15pt}
\begin{align*}
\text{s.t.} \quad \quad (\ref{const:first})-(\ref{const:last}). 
\end{align*}

The optimization problem (1)-(18) is an ILP since all the decision variables are binary, and the relation between them in all the constraints are linear. 

\subsection{NP-hardness}
\label{subsec:complexity}

We show that the ILP at hand is NP-hard. There are several aspects that need to be considered in (1)-(18) in terms of complexity. However, due to space limitations we focus only on proving that demand placement in an OEN is computationally expensive. 
\begin{lemma}
Demand placement in an edge cloud node with limiting CPU resources is NP-hard.
\end{lemma}
\begin{proof}
The NP-hardness of this aspect is proven by reducing a 0-1 Knapsack Problem (KP) to it. There is a set $\mathcal{X}$ with $n$ items (in our case, a set of $n$ demands). Each item has value $v_i$ (each demand has utility $U_i$) and weight $w_i$ (CPU requirement $p_i$).\footnote{Note that in this proof we have simplified the notation by reducing the number of indices for utility and CPU requirement from $\{j,i,l\}$ to $i$.} In a 0-1 KP the goal is to select a set of possible items whose weights do not exceed the total weight of the knapsack while maximizing the total value. In our case, that would correspond to placing a set of demands whose total CPU requirements do not exceed the CPU capacity of the edge cloud while maximizing the total utility.
If a given demand $i$ is placed on the edge cloud, the corresponding $p_i$ is deducted from the CPU capacity of the edge cloud. On the other hand, it is well known that the 0-1 KP is NP-hard~\cite{Kellerer}. Therefore, maximizing the total utility on an edge cloud by deciding which demands to place on it is NP-hard.  
\end{proof}

In a similar vein, it can be proven that placing Edge Apps is a two-dimensional KP (in storage and CPU). The same holds when proving UPF placement (in CPU) is a KP. Due to space limitations, we skip these proofs. Finally, combining the previous claims and Lemma~1, we have the following:
\begin{result}
The optimization problem (1)-(18) is NP-hard. 
\end{result}

%% file: heuristic.tex
\section{RanGr}
\label{sec:rangr}

\SetStartEndCondition{ }{}{}
\SetKwProg{Fn}{def}{\string:}{}
\SetKwFunction{FMain}{Main}

\SetKw{KwTo}{in}
\SetKwFor{For}{for}{\string:}{}
\SetKwIF{If}{ElseIf}{Else}{if}{:}{elif}{else:}{}
\SetKwFor{While}{while}{:}{fintq}
\newcommand{\forcond}{$i$ \KwTo\Range{$n$}}
\newcommand{\forin}[3]{\For{#1 \KwTo #2}{#3}}

\AlgoDontDisplayBlockMarkers
\SetAlgoNoLine

\begin{algorithm}
	\label{algo:rangr}
	\caption{RanGr}
	\DontPrintSemicolon
	\KwIn{\begin{tabular}{ll}
			\textit{D} 							& List of input demands \\
			\textit{BSList} 					& List of BSs in the topology\\
			\textit{OENList} 					& List of OENs and their parameters \\
			\textit{EEN} 						& EEN and its parameters \\
	\end{tabular}}

	\KwOut{Demands' anchoring and placement scheme}
	
	\SetKwFunction{sort}{sort}
	\SetKwFunction{sortPaths}{getAndSortPaths}	
	\SetKwFunction{anchorAndPlaceDemands}{anchorAndPlaceDemands}
	\SetKwFunction{undoAnchorAndPlacement}{undoAnchorAndPlacement}
	\SetKwFunction{getRejectedDemands}{getRejectedDemands}
	\SetKwFunction{calculateProfit}{calculateProfit}
	\SetKwFunction{append}{.append}
	\SetKwFunction{getTotalRequestedCPU}{getTotalRequestedCPU}
	\SetKwFunction{rankD}{rankDemand}
	\SetKwFunction{initList}{initEmptyList}
	\SetKwFunction{rankOEN}{rankOEN}
	\SetKwFunction{getOEN}{getTopOEN}
	\SetKwFunction{canPlace}{canPlaceInEdge}
	\SetKwFunction{placeDemand}{placeInEdge}
	\SetKwFunction{canAnchor}{canAnchorInOEN}
	\SetKwFunction{anchorDemand}{anchorInOEN}
	\SetKwFunction{reject}{rejectDemand}
	\SetKwFunction{appendAnchored}{anchoredDemands.append}

	\BlankLine
	\Fn{\FMain{}}{
		
		\sortPaths{BSList, OENList, EEN}\;\label{alg:paths}
		
		\forin{d}{D}{ \label{alg:startRank}
			\textit{d\textsubscript{rank}} $\gets$ \rankD{d\textsubscript{U}, d\textsubscript{C}, d\textsubscript{CPU}}\;
		}
		\textit{D\textsubscript{sorted}} $\gets$ \sort{D, key $\gets$ d\textsubscript{rank}}\;\label{alg:endRank}
		
		\anchorAndPlaceDemands{D\textsubscript{sorted}, OENList, EEN}\;\label{alg:anchorAndPlace}
		\textit{D\textsuperscript{R}} $\gets$ \getRejectedDemands{}\;\label{alg:startReallocation}
		\forin{OEN}{OENList}{
			\If{OEN\textsubscript{Profit} $\le$ OEN\textsubscript{ON\_cost}}{
				\textit{D\textsuperscript{R}\textsubscript{OEN}} $\gets$ \undoAnchorAndPlacement{}\;
				\textit{D\textsuperscript{R}}\append{D\textsuperscript{R}\textsubscript{OEN}}\;
			}
		}
		\textit{D\textsuperscript{R}\textsubscript{sorted}} $\gets$ \sort{D\textsuperscript{R}, key $\gets$ d\textsubscript{rank}}\;
		\anchorAndPlaceDemands{D\textsuperscript{R}\textsubscript{sorted}}\;\label{alg:endReallocation}
	}
	
	\BlankLine
	
	\Fn{\anchorAndPlaceDemands{D\textsubscript{sorted}, OENList, EEN}}{\label{startFunction}
		\initList(\textit{anchoredDemands})\;
		\forin{d}{D\textsubscript{sorted}}{\label{alg:startAnchor}
			\textit{d\textsubscript{anchored}} $\gets$ \textit{False}\;
			\forin{OEN}{OENList}{\label{alg:startRankOne}
				\textit{OEN\textsubscript{R\textsubscript{P}}}, \textit{OEN\textsubscript{R\textsubscript{A}}} $\gets$ \rankOEN{OEN}\;
			}\label{alg:endRankOne}
			\textit{rankedOENs} $\gets$ \sort{OENList, key $\gets$ OEN\textsubscript{R\textsubscript{P}}}\;\label{alg:sortOENOne}
			\textit{topOEN} $\gets$ \textit{rankedOENs[0]}\;
			\If{\canPlace{topOEN}}{
				\anchorDemand{d, topOEN}\;
				appendAnchored{d} \;
				\textit{d\textsubscript{edge}} $\gets$ \textit{topOEN}\;
				\textit{d\textsubscript{anchored}} $\gets$ \textit{True}\;
			}
			\Else{
				\textit{rankedOENs} $\gets$ \sort{OENList, key $\gets$ OEN\textsubscript{R\textsubscript{A}}}\;\label{alg:sortOENTwo}
				\forin{OEN}{rankedOENs}{\label{alg:startRankTwo}
					\If{\canAnchor{OEN}}{
						\anchorDemand{d, OEN}\;
						\textit{d\textsubscript{anchored}} $\gets$ \textit{True}\;
						\textbf{break}\;\label{alg:endRankTwo}
					}
				}
			}
			\If{\textit{d\textsubscript{anchored}} is \textbf{False}}{
				\reject{d} \;
			}
		}\label{alg:endAnchor}
		\forin{d}{anchoredDemands}{\label{alg:startPlace}
			\If{\canPlace{d\textsubscript{edge}}}{
				\placeDemand{d, d\textsubscript{edge}} \;
			}
			\ElseIf{\canPlace{EEN}}{
				\placeDemand{d, EEN} \;
			}
			\Else{
				\reject{d} \;\label{alg:endPlace}
			}
		}
	}\label{endFunction}
\end{algorithm}

In Section~\ref{subsec:complexity}, we showed that the joint UPF and demand placement and routing problem is a variation of the d-dimensional Knapsack Problem (d-KP)~\cite{Kellerer} and the Multiple Knapsacks Problem (MKP)~\cite{Kellerer}.
As such, our problem is shown to be NP-Hard, and finding the optimal solution with a solver is computationally very expensive.
Therefore, to overcome this inherent limitation of any variation of the KP, we design and implement our own ranked greedy algorithm called \emph{RanGr}.
As will be shown in Section~\ref{sec:results}, \emph{RanGr} finds near-optimal solutions in a very short time.

Algorithm~$\ref{algo:rangr}$ provides a detailed overview of the \emph{RanGr} execution. For enhanced clarity, we have separated the definitions of the \texttt{Main} and \texttt{anchorAndPlaceDemands} procedures.
Initially, the algorithm calculates paths from each BS as the source to every OEN and EEN as the destinations, based on the given network topology.
The function called in line~$\ref{alg:paths}$ does not calculate only the shortest path, but all the paths between two nodes that are shorter than a predefined \emph{cutoff} value.
This threshold on the maximum number of hops in the path is used as a stopping condition for the path calculation, and is based on the fact that routing demands via longer paths may violate their latency constraint.
In analogy with the ILP's link-path formulation, this is the same function that calculates the list of paths which are then given as input to the ILP~$(\ref{obj})-(\ref{const:last})$, with the only difference being that in \emph{RanGr} the paths are also sorted based on the number of hops.

\emph{RanGr} is designed as a \emph{primal} greedy heuristic, meaning that the demands are ranked and sorted in descending order of their rank (lines~$\ref{alg:startRank}$-$\ref{alg:endRank}$) before executing the anchoring and placement procedure in line~$\ref{alg:anchorAndPlace}$.
More details on the ranking formula and the anchoring and placement algorithm are given in the following subsections.
In lines~$\ref{alg:startReallocation}$-$\ref{alg:endReallocation}$ a \emph{``reallocation''} process takes place in the following order.
First, a list of the rejected demands is created.
Next, \emph{RanGr} evaluates if the OENs have become profitable.
Therefore, if the profit from demands anchored or placed on the OEN is not greater than its operational costs, the algorithm proceeds to undo the anchoring/placement of the demands from that OEN, and appends them to the list of rejected demands.
Lastly, the list is sorted based on the rank of the demands, and the anchoring and placement algorithm is executed again for this subset of the initial demands.
This step marks the end of \emph{RanGr}'s execution, and the final UPF and EA deployment schemes, demands' anchoring and placement schemes and the selected routes are obtained.

\subsection{Demand Ranking}
\label{subsec:ranking}
As in other greedy heuristics for the KP~\cite{Kellerer}, it is important to define a ranking criteria that efficiently captures the ``attractiveness'' of anchoring/placing a demand, in order to obtain high-quality solutions that maximize the operator's profit.
For the classic KP, the \emph{efficiency} of an item is defined as the \emph{profit to weight ratio}, where weight is the only constraint that determines whether an item can be placed into the knapsack.
In our problem, although there are many constraints on the resources at hand, most of them serve the purpose of identifying the subset of OENs where the demand can be anchored/placed on (e.g., the latency and bandwidth constraints).
Therefore, CPU and storage represent the only resources counterpart to the weight in the classic KP.
As the latter is related to the EA that will serve the demands and not to the demand itself directly, for the sake of simplicity, it is omitted from the rank calculation. Summarizing, we have the following:
\begin{definition}
The rank/efficiency of a demand is defined as
\begin{align}
	\textit{d}_{\textit{rank}}= \left( \frac{\textit{d}_\textit{U}}{\textit{d}_{\textit{CPU}}}, -(\textit{d}_\textit{U} - \textit{d}_\textit{C}) \right),
\end{align}
where $\textit{d}_{\textit{U}}$, $\textit{d}_{\textit{C}}$ and $\textit{d}_{\textit{CPU}}$ represent the utility, offloading cost, and CPU requirement of the demand, respectively.
\end{definition}
To prevent any unintentional bias caused by varying ranges among the values, all data are normalized.
The rank of the demand is defined as a tuple, where the ratio of utility to CPU requirement takes precedence. In cases when this ratio is identical for more than one demand, the difference between its utility and cost (i.e., the profit margin) is the deciding factor.
Following this ranking methodology, we compile a list where demands that are more \emph{profitable} when placed on OENs, and that have a lower profit margin when offloaded to the EEN, are ranked higher.

At a first glance, it may seem that the chosen ranking formula is too simple for such a complex problem.
Nonetheless, we experimented with other formulas that take into account the relationship between utility, cost, and CPU resources in a single equation rather than a tuple, but the quality of the solution did not improve, and in some cases it even degraded.

\subsection{Anchoring and Placement Procedure}
\label{subsec:anchoring}
After the demands are ranked and sorted, in line~$\ref{alg:anchorAndPlace}$, the anchoring and placement procedure (lines~$\ref{startFunction}$-$\ref{endFunction}$) is executed. 
This procedure implements a best-fit algorithm, which identifies the most suitable edge node for each demand being evaluated.
The objective of maximizing the operator's profit is translated in this procedure to: i) maximize the number of accepted (i.e., anchored) demands, ii) maximize the number of highly-profitable demands placed on OENs, and iii) minimize the number of OENs in operation mode.
Anchoring demands on an OEN means that UPF instances need to be initialized on the OENs, and CPU resources have to be allocated to these UPFs. 
At the same time, OENs' CPU resources constrain the placement of demands onto these OENs.
For this reason, the demand anchoring and placement procedures are separated from each other and executed sequentially.

\emph{RanGr} tries first to anchor as many demands as possible (lines~$\ref{alg:startAnchor}$-$\ref{alg:endAnchor}$).
For each demand being evaluated, it ranks the OENs by calculating:
\begin{itemize}
	\item \textit{OEN}\textsubscript{\textit{R}\textsubscript{\textit{A}}} -- the remaining CPU resources at OEN after anchoring the demand, defined as:
	\begin{align}
		\textit{OEN}\textsubscript{\textit{R}\textsubscript{\textit{A}}} \hspace{-2pt}= \hspace{-2pt}\textit{OEN}\textsubscript{\textit{CPU$^\prime$}} \hspace{-2pt}- \hspace{-2pt}\textit{UPF}\textsubscript{\textit{upscaleCPU}},
	\end{align}
	where \textit{OEN}\textsubscript{\textit{CPU$^\prime$}} is the amount of available CPU resources, and \textit{UPF}\textsubscript{\textit{upscaleCPU}} represents the CPU resources to be allocated to the UPF in case that anchoring the demand would lead to a UPF upscale.
	
	\item \textit{OEN}\textsubscript{\textit{R}\textsubscript{\textit{P}}} -- the remaining CPU resources in the OEN after anchoring and placing the demand, defined as:
	\begin{align}
		\hspace{-2pt}\textit{OEN}\textsubscript{\textit{R}\textsubscript{\textit{P}}} \hspace{-2pt}= \hspace{-2pt}\textit{OEN}\textsubscript{\textit{CPU$^{\prime\prime}$}} \hspace{-2pt}- \hspace{-2pt}\textit{UPF}\textsubscript{\textit{upscaleCPU}} \hspace{-2pt}- \hspace{-2pt}\textit{EA}_{\textit{CPU}} \hspace{-2pt}- \hspace{-2pt}\textit{d}_{\textit{CPU}},
	\end{align}
	where \textit{OEN}\textsubscript{\textit{CPU$^{\prime\prime}$}} is the amount of available CPU resources assuming that previous demands anchored on this OEN are also placed there, $\textit{EA}_{\textit{CPU}}$ is the CPU requirement of the EA that would serve the demand (considered only in the first occurrence of the EA in the given OEN), and $\textit{d}_{\textit{CPU}}$ is the demand's CPU requirement.
\end{itemize}

Since the demands can be placed either on the same OEN where they are anchored, or on EEN, we can think of the anchoring procedure as a non-binding placement procedure.
After OENs are ranked, they are sorted by the value of OEN\textsubscript{R\textsubscript{P}}, in decreasing order.
The viability of the top-ranked OEN to place the demand is then evaluated.
If placing the demand would not violate any of the constraints, \emph{RanGr} proceeds to anchor the demand on the OEN and save the placement scheme for later.
If placing would not be possible, then the OENs are again sorted in decreasing order, but this time by the value of OEN\textsubscript{R\textsubscript{A}}.
The algorithm iterates over the list of sorted OENs for anchoring, and once it finds a suitable OEN, it anchors the demand and breaks the loop.
In cases where a suitable OEN could not be found, the demand is rejected.

Lastly, \emph{RanGr} executes the placement procedure (lines~$\ref{alg:startPlace}$-$\ref{alg:endPlace}$) by iterating through the demands that were successfully anchored.
Here, \emph{RanGr} evaluates again if it would be possible to place the demand on the same OEN where it is anchored on.
The reason why this check is performed again is due to the scenarios where as \emph{RanGr} evaluates more demands, CPU resources end up being used to initialize more UPFs, and the assumed placement schemes for some of the demands may not be valid anymore.
If the placement scheme is still valid, \emph{RanGr} proceeds to place the demands; otherwise, it checks if the demand can be offloaded to the EEN (while still being anchored on the original OEN).
If that is not the case (e.g., the latency requirement cannot be satisfied), the demand is rejected.

\subsection{Complexity Analysis of RanGr}
\label{subsec:rangr_analysis}
To analyze the worst-case complexity of \emph{RanGr}, we assume $\vert E \vert$ to be the number of OENs and $\vert D \vert$ the number of input demands.
We exclude the complexity of the \texttt{getAndSortPaths} function because that will be executed only once.
Further, we will evaluate separately the two main operations executed inside the \texttt{Main} function, before deriving the overall complexity of \emph{RanGr}.

\subsubsection{Demands ranking and sorting procedure}
Ranking is an operation executed individually for each demand, with a constant-time complexity, $O(1)$. Hence, ranking takes $O(\vert D \vert)$.
Following, the demands are sorted (using  Python’s \texttt{sorted()} -- a variant of merge sort) based on their calculated ranks. 
Merge sort is known to have a complexity of $O(\vert D\vert\log\vert D\vert)$.
Due to this linearithmic complexity, sorting dominates the \textit{linear} ranking complexity.
Thus, the overall complexity for ranking and sorting is $O(\vert D\vert\log\vert D\vert)$.

\subsubsection{Demands anchoring and placing procedure}
This anchoring part of the algorithm involves several nested operations executed while iterating over the sorted list of demands.
Specifically, each iteration (line~\ref{alg:startAnchor}) processes one demand from the list, with $\vert D \vert$ total iterations.
A detailed analysis of the operations within this loop yields: \\
  \noindent \textit{1) Nested loops over OENs}:
	The algorithm contains two distinct loops iterating over the list of OENs, found at lines~$\ref{alg:startRankOne}$-$\ref{alg:endRankOne}$ and $\ref{alg:startRankTwo}$-$\ref{alg:endRankTwo}$.
	Each operation inside these loops executes in $O(1)$.
	Hence, each loop contributes $O(\vert E \vert)$ for a combined complexity of $O(2 \times \vert E \vert)$.
	As constant factors are omitted for simplification, it can be simplified to $O(\vert E \vert)$.\\
      \noindent \textit{2) Sorting operations}:
	Within this part of the algorithm, there are two occurrences of sorting the list of OENs based on different criteria.
	Considering the complexity of merge sort, and the simplification by omitting the constant factors, the complexity of this part is $O(\vert E \vert \log \vert E \vert)$. 

Combining these elements, sorting operations within each iteration dominate the complexity.
Thus, for $\vert D \vert$ iterations, the total complexity associated with anchoring part of the algorithm becomes $O(\vert D\vert\times\vert E\vert\log\vert E\vert)$.
The placing procedure, spanning lines~$\ref{alg:startPlace}$-$\ref{alg:endPlace}$, iterates over the list of demands that have been successfully anchored.
As this number can potentially reach $\vert D\vert$, the placing procedure exhibits a complexity of $O(\vert D\vert)$.
The anchoring procedure clearly dominates in terms of complexity due to the inclusion of the logarithmic and multiplicative factors involving $\vert E\vert$.
Thus, the worst-case complexity for the \texttt{anchorAndPlaceDemands} procedure remains $O(\vert D\vert\times\vert E\vert\log\vert E\vert)$. 

Finally, deriving the worst-case complexity of \emph{RanGr} based on the analysis of the two main parts of the algorithm, we obtain $O(\vert D\vert\log\vert D\vert + \vert D\vert\times\vert E\vert\log\vert E\vert)$.

%% file: results.tex
\section{Performance Evaluation}
\label{sec:results}
We implement \emph{RanGr} in Python and compare its performance with the ILP solved with Gurobi~\cite{gurobi}, and two benchmark algorithms.
The evaluation setup consists of a Virtual Machine (VM), to which we allocate $20$\:CPU cores and $64$\:GB of RAM.
In the following, we discuss in detail the evaluation scenarios and the obtained results.

\begin{figure}[t]	
	\centering
	\subfloat[Small]{
		\includegraphics[scale=0.85]{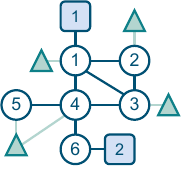}}
	\hfill
	\subfloat[Medium]{
		\includegraphics[scale=0.85]{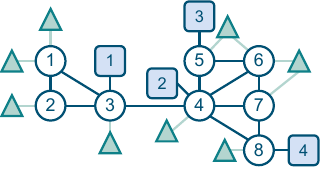}}
	\hfill
	\subfloat[Large]{
		\includegraphics[scale=0.85]{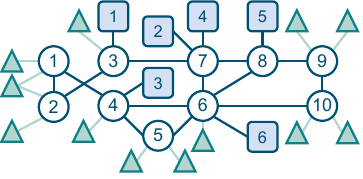}}
	\vfill
	\caption{Edge network topologies considered for the evaluation.
	The configuration of the number of BSs (triangles), switches (circles), and OENs (squares) for each topology is: a) $(4, 6, 2)$, b) $(8, 8, 4)$, and c) $(12, 10, 6)$.
	The same amount of resources is allocated to each OEN and link in each topology.
	}
	\label{fig:topos}
\vspace{-10pt}
\end{figure}

\subsection{Benchmarks}
With reference to algorithm evaluations performed in the literature for similar problems~\cite{Li2021, Li2022},~\cite{PupoUPFChain2022, Farhadi2019, Xiang2019}, we have implemented two benchmark algorithms to compare to \emph{RanGr}.
These are: i) Greedy, and ii) Top-K.
Instead of \emph{RanGr}'s ranking mechanism, these algorithms rank the demands based on their utility values.
Moreover, \emph{RanGr} dynamically initializes OENs and UPF clusters. 
Greedy and Top-K determine these aspects in the beginning, based on the CPU and bandwidth requirements of the input demands.
Lastly, \emph{RanGr} employs a best-fit strategy when selecting on which OEN to place the demand, while the benchmark algorithms utilize a first-fit approach, where demands are placed on the first viable OEN.

\begin{figure*}[t!]	
	\centering
	\subfloat[Total profit]{
		\label{fig:profit}
		\includegraphics{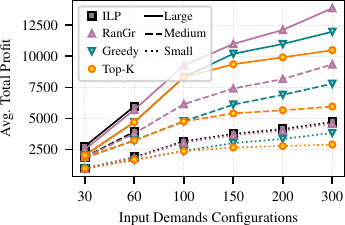}}
	\subfloat[CPU utilization (large topology)]{
		\label{fig:cpu}
		\includegraphics{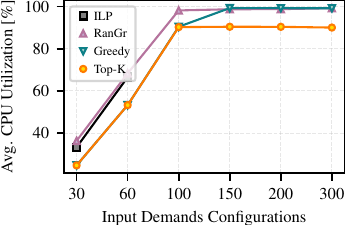}}
	\subfloat[Demands placed on OENs]{
		\label{fig:demands}
		\includegraphics{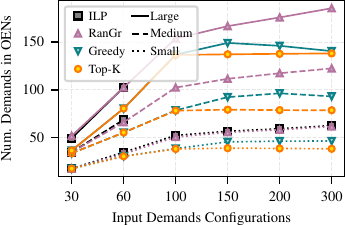}}
	\caption{Performance of the algorithms for different configurations of input demands, in terms of a) total profit, b) OEN CPU utilization, and c) number of demands placed on OENs. Results for the small, medium and large topologies are shown in dotted, dashed, and solid lines, respectively.
	}
	\label{fig:results}
	\vspace{-10pt}
\end{figure*}
\subsection{Evaluation Scenarios}

\begin{table}
	\centering
	\caption{Parameters used for input demands generation}
	\begin{tabular}{|c|c|}
		\hline
		Demand parameter & Value                   \\ \hline
		EA type                   & {[}$1, 2, 3, 4, 5${]}              \\ \hline
		CPU req.                  & {[}$750, 1000, 1250, 1500${]}\:mCPU \\ \hline
		Bandwidth req.            & {[}$30, 40, 50, 60${]}\:Mbps        \\ \hline
		Delay budget              & {[}$3, 4, 5${]}\:ms                 \\ \hline
		Cost of EEN offloading    & {[}$40, 50, 60, 70${]}\:units       \\ \hline
		Utility                   & {[}$44-91${]}\:units                \\ \hline
	\end{tabular}
\label{tab:demands}
\vspace{-5pt}
\end{table}

We compare the performance of \emph{RanGr} for three different topologies of different sizes: i) small, ii) medium, and iii) large, as illustrated in Figure~\ref{fig:topos}.
Each Operator Edge Node (OEN) is assumed to have $32$\:CPU cores ($32000$\:mCPU) and $250$\:GB of storage, while the links have a capacity of $10$\:Gbps.
Moreover, we assume a switching on cost of $200$\:units for OENs, $1$\:ms latency for the transport links, and $1.5$\:ms latency for the links connecting OENs to EEN.
The sets of input demands are generated as a percentage of the total OEN CPU resources that they require.
Thus, sets with demands that in total have CPU requirements amounting to $[30, 60, 100, 150, 200, 300]$\% of the capacity of OENs are generated.
For each scenario, we generate $100$ different inputs, where the parameters of a demand are picked randomly from a pre-defined set, and each demand is assigned randomly to a BS.
Due to the high execution time, ILP results are available only for the small topology, and for the medium and large topologies, results are obtained only for the lightly-loaded scenarios.
The demand parameters are summarized in Table~\ref{tab:demands}.

\subsection{Results}

\subsubsection{Profit}
In Fig.~\ref{fig:profit} shows the average total operator's profit for different topologies and demands.
\emph{RanGr} exhibits a near-optimal performance consistently, deviating only $\sim$\:$3\%$ from the optimal solution sound by solving the ILP with Gurobi.
Greedy and Top-K, on the other hand, perform good only for the lightly-loaded scenarios, and as the number of demands increases, their performance degrades, deviating on average $\sim19\%$ and $\sim33\%$ from the optimal solution.
In highly-loaded scenarios, \emph{RanGr} consistently outperforms Greedy, with the latter deviating by on average $\sim18\%$ and $\sim10\%$ from \emph{RanGr}'s solution, for the medium and large topologies, respectively.
The total profit obtained by using Top-K is on average $\sim29\%$ lower than \emph{RanGr} for the medium topology, and $\sim17\%$ for the large topology.
The superior performance of \emph{RanGr} can be attributed to its ranking mechanism, and the best-fit strategy taken for anchoring and placing the demands in the edge infrastructure.

\subsubsection{CPU Utilization and Demands in OENs}
Fig.~\ref{fig:cpu} shows the average CPU utilization (considering all OENs) for the large topology.
\emph{RanGr} again performs very good, reaching $\sim98\%$ CPU utilization as the demands increase.
However, it is important to also evaluate how this resources are distributed.
Fig.~\ref{fig:demands}, shows the number of demands placed on the OENs for all the topologies.
In the case of \emph{RanGr}, this number increases consistently as the input size increases, proving that the ranking operation can efficiently capture the value of the demands.
In terms of CPU utilization, Top-K could only reach up to $\sim90\%$, for all the overloaded scenarios.
By investigating the average number of demands it has placed in OENs, we see that it is almost static for all the highly-loaded scenarios.
The reason is that Top-K considers only $K$ demands for placement, leading to inefficient CPU utilization.
Lastly, Greedy also reaches $\sim98\%$ CPU utilization, but in Fig.~\ref{fig:demands} we see that it performs bad in allocating these resources to the demands.
Specifically, this is a result of the inefficient ranking mechanism, which allocates \emph{``heavy''} demands, and saturating the available resources as a result.

\subsubsection{Execution Time}
Table~\ref{tab:time} shows the average execution times in milliseconds for \emph{RanGr}, Greedy, and Top-K, for different topologies and number of demands.
Generally, all three algorithms perform good, finishing within a few milliseconds. 
However, there are a few interesting points.
\emph{RanGr}'s execution time grows linearithmically with the increase of the size of the demands set, and this is observed for all three topologies.
The results are in line with the complexity analysis in Section~\ref{subsec:rangr_analysis}, where we showed that for a high numbers of demands, the worst-case complexity is dominated by the ranking procedure.
Greedy seems to finish the execution faster than \emph{RanGr} for scenarios where the system is not overloaded.
This changes when the number of demands is considerably above what the OEN network can host, where Greedy takes almost double the time compared to \emph{RanGr}.
This is as a result of the additional operations for initializing OENs and UPF clusters, executed in the beginning. 

Top-K exhibits the same behavior as Greedy for the first three scenarios, and for all the topologies.
However, increasing the size of demands set has no impact on the execution time, because the number $K$ of demands to process is calculated only based on the amount of resources in the OENs.

\begin{table}
	\centering
	\caption{Average algorithm execution time in milliseconds}
\begin{tabular}{|c|c|cccccc|}
	\hline
	\multirow{2}{*}{Topology} & \multirow{2}{*}{Algorithm} & \multicolumn{6}{c|}{\% of OENs' CPU needed for demands}                                                                                         \\ \cline{3-8} 
	&                            & \multicolumn{1}{c|}{30}  & \multicolumn{1}{c|}{60}  & \multicolumn{1}{c|}{100} & \multicolumn{1}{c|}{150}  & \multicolumn{1}{c|}{200}  & 300  \\ \hline
	\multirow{3}{*}{Small}    & RanGr                      & \multicolumn{1}{c|}{0.6} & \multicolumn{1}{c|}{1.1} & \multicolumn{1}{c|}{1.9} & \multicolumn{1}{c|}{2.9}  & \multicolumn{1}{c|}{4.0}  & 6.3  \\ \cline{2-8} 
	& Greedy                     & \multicolumn{1}{c|}{0.4} & \multicolumn{1}{c|}{0.8} & \multicolumn{1}{c|}{1.7} & \multicolumn{1}{c|}{3.0}  & \multicolumn{1}{c|}{4.8}  & 7.6  \\ \cline{2-8} 
	& Top-K                       & \multicolumn{1}{c|}{0.4} & \multicolumn{1}{c|}{0.8} & \multicolumn{1}{c|}{1.7} & \multicolumn{1}{c|}{1.7}  & \multicolumn{1}{c|}{1.7}  & 1.7  \\ \hline
	\multirow{3}{*}{Medium}   & RanGr                      & \multicolumn{1}{c|}{1.4} & \multicolumn{1}{c|}{2.6} & \multicolumn{1}{c|}{4.5} & \multicolumn{1}{c|}{7.1}  & \multicolumn{1}{c|}{9.6}  & 16.0 \\ \cline{2-8} 
	& Greedy                     & \multicolumn{1}{c|}{0.9} & \multicolumn{1}{c|}{2.3} & \multicolumn{1}{c|}{4.9} & \multicolumn{1}{c|}{10.1} & \multicolumn{1}{c|}{16.8} & 30.4 \\ \cline{2-8} 
	& Top-K                       & \multicolumn{1}{c|}{0.8} & \multicolumn{1}{c|}{2.2} & \multicolumn{1}{c|}{4.9} & \multicolumn{1}{c|}{4.9}  & \multicolumn{1}{c|}{4.9}  & 5.0  \\ \hline
	\multirow{3}{*}{Large}      & RanGr                      & \multicolumn{1}{c|}{2.4} & \multicolumn{1}{c|}{4.5} & \multicolumn{1}{c|}{7.5} & \multicolumn{1}{c|}{12.5} & \multicolumn{1}{c|}{17.8} & 25.7 \\ \cline{2-8} 
	& Greedy                     & \multicolumn{1}{c|}{1.2} & \multicolumn{1}{c|}{3.5} & \multicolumn{1}{c|}{8.5} & \multicolumn{1}{c|}{17.5} & \multicolumn{1}{c|}{27.0} & 51.1 \\ \cline{2-8} 
	& Top-K                       & \multicolumn{1}{c|}{1.2} & \multicolumn{1}{c|}{3.5} & \multicolumn{1}{c|}{8.2} & \multicolumn{1}{c|}{8.4}  & \multicolumn{1}{c|}{8.3}  & 8.7  \\ \hline
\end{tabular}
\label{tab:time}
\end{table}

%% file: related_work.tex
\section{Related Work}
\label{sec:related_work}
Placement problems find application in many different areas of mobile core networks.
The edge server placement problem has been investigated for different objectives and with different considerations~\cite{Lee2019, Li2022, Wang2019, Li2018, Xu2019}.
From a profit perspective, the authors in~\cite{Lee2019} minimize the number of MEC locations and model the relation between BSs and MECs, while the authors in~\cite{Li2022} additionally consider UPF deployment when dimensioning the edge infrastructure and formulate an objective that maximizes the operator's profit.
The authors in~\cite{Wang2019} optimize the placement of edge servers in a BS topology, while minimizing the access delay.
Energy-aware~\cite{Li2018} and load-aware~\cite{Xu2019} deployment procedures are also investigated.
A joint approach to the edge server and service placement problem is given in~\cite{Zhang2022}.
However, these works assume a network planning approach, trying to optimally dimension the edge infrastructure for any future services that might be deployed.
Here, we formulate the problem for an already deployed edge infrastructure, and consider aspects related to service and UPF placement, as well as demand routing.

Leyva-Pupo \emph{et al.} formulate the problems of minimizing capital and operating expenditures when deploying UPFs~\cite{PupoUPF2019} and UPF chains~\cite{PupoUPFChain2022} in the 5G infrastructure.
In another research work, the authors take a more holistic approach for jointly dimensioning the edge infrastructure and deploying UPFs.
A multi-objective framework is presented in~\cite{PupoJoint2019} which solves the multi-objective optimization problem of minimizing the number of edge nodes, UPFs, and relocations.

Service placement and scheduling for edge applications is investigated in~\cite{Farhadi2019}, while the problem of jointly placing services in edge servers located in BSs, and routing user requests in the access network is investigated in~\cite{Poularakis2020}. 
Lastly, the authors in~\cite{Xiang2019} propose a model to jointly slice the network and compute resources available at the edge, while minimizing the overall user-experienced latency.

Therefore, to the best of our knowledge, this is the first work that models and formulates the problem of jointly orchestrating the deployment of UPFs, Edge Applications (EAs), and demands on the edge infrastructure, with the objective of maximizing TNO's profit.

%% file: conclusions.tex
\section{Conclusion}
\label{sec:conclusions}
In this paper, we considered the problem of jointly placing UPFs and edge applications as well as routing the demands, encompassing the entire cellular network.
The goal was to maximize the operator's profit. 
The different traffic requirements of the tasks are taken into account, including the tolerable maximum latency of each demand, as well as the limitations across different types of resources in the network.
The resulting optimization problem is NP-hard, and we resort to using a heuristic, \emph{RanGr}.
Results showed that the performance of \emph{RanGr} is near-optimal and considerably better than the benchmark.
In the future, multi-objective formulations that consider the energy consumption of edge nodes and scenarios with dynamic input demands will be considered.